# Biological and Statistical Heterogeneity in Malaria Transmission

Brian G. Williams[1] and Christopher Dye[2]

1. South African Centre for Epidemiological Modelling and Analysis (SACEMA), Stellenbosch, South Africa
2. Office of Health Information, HIV/AIDS, Tuberculosis, Malaria & Neglected Tropical Diseases, World Health Organization, Geneva, Switzerland
   Correspondence to BGW at briangerardwilliams@gmail.com

## Abstract

In a study of the heterogeneity in malaria infection rates among children, Smith *et al.*[1] fitted several mathematical models to data from community studies in Africa. They concluded that 20% of children receive 80% of infections, that infections last about six months on average, that children who clear infections are not immune to new infections, and that the sensitivity and specificity of microscopy for the detection of malaria parasites are 95.8% and 88.4%, respectively. These findings would have important implications for disease control, but we show here that the statistical analysis is unsound and that the data do not support their conclusions.

## Introduction

Heterogeneity in transmission plays a crucial role in the emergence and spread of infectious diseases[2] including malaria.[1,3] Smith *et al.*[1] assembled data from 91 studies in different countries in Africa on the entomological inoculation rate and the parasite rate for malaria in children up to the age of fifteen years. They fitted the following models to the data: 1) log-linear, not based explicitly on the biology of malaria; 2) SIS, a dynamical model in which susceptible children are infected but return to the susceptible state if they clear the infection; 3) SI°S: which allows for super-infection; 4) ∫SI°S: which includes heterogeneous infection rates. They also fitted models in which recovered children become immune. To choose among the models they used the Akaike Information Criterion (AIC).

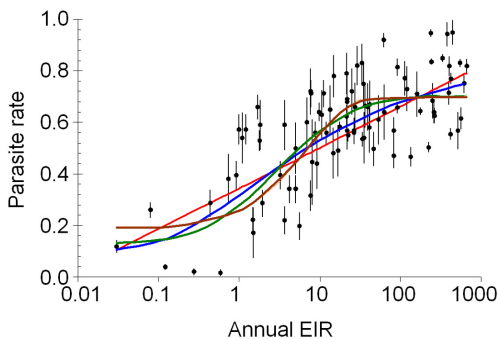

Figure 1. The relationship between the entomological inoculation rate (EIR) and the parasite rate.[1] The points are the data, the error bars are normal approximations to the 95% binomial confidence limits, the lines give four models: red log-linear; blue ∫SI°S; green SIS; brown SI°S.

## Analysis

The data, as given by the authors, are plotted in Figure 1 with 95% confidence limits corresponding to the binomial sampling errors in each study together with the predictions of the four models. The author's estimates of the change in AIC, $\Delta A_1$, are given in Table 1. Because the AIC for the SIS and SI°S models are greater than for the ∫SI°S model, which includes super-infection and heterogeneity in susceptibility to infection, they conclude that super-infection and heterogeneity are both important. Because the inclusion of immunity does not significantly change the predictions of the model (data not shown) they conclude that immunity is not important.

It is well known that the absolute value of the AIC cannot be used to judge the goodness of fit directly and changes in the AIC from the best-fit model, $\Delta A$, are used to decide if other models can be rejected in comparison with the best fit model in which case the data give significantly more support for the best fit model as compared to the others. By definition

$$e^{-\Delta A/2} = \frac{L_2}{L_1} e^{\Delta n} \qquad 1$$

so that $e^{-\Delta A/2}$ gives the likelihood ratio for two models penalizing models with more parameters.[4] If the author's best fit model, ∫SI°S, is statistically a good fit to the data, then the likelihood ratios for their other models compared to the best fit model are all less than $10^{-41}$.

Table 1. The Akaike Information Criterion calculated using binomial sampling errors[1] and using the residual sum-of-squares for the four models described in the text. $\Delta A_1$ gives the estimates, published by Smith *et al.*[1], of the difference in the AIC for each model from ∫SI°S, the best fit model. Large values of $\Delta A$ are taken to indicate support for the best fit model as compared to the comparison model. $LR_1$ gives the likelihood ratio for the alternative model relative to the best fit model. *Parameters* gives number of parameters for each model; $A_2$ gives the AIC estimated in this paper from the residual sums-of-squares; $\Delta A_2$ the difference in the AIC for each model from the best fit model.; $LR_2$ gives the likelihood ratio for the alternative model relative to the best fit model as estimated by the present authors.

|  | ∫SI°S | Log-linear | SIS | SI°S or SI°RS |
|---|---|---|---|---|
| $\Delta A_1$ | 0 | 185 | 213 | 557 or 612 |
| $LR_1$ | 1 | $9\times10^{-42}$ | $2\times10^{-47}$ | $4\times10^{-122}$ |
| Parameters | 6 | 4 | 5 | 5 |
| $A_2$ | 96.02 | 97.59 | 96.06 | 100.98 |
| $\Delta A_2$ | 0 | 1.59 | 0.06 | 4.98 |
| $LR_2$ | 1 | 0.061 | 0.357 | 0.031 |

The mistake is to use the binomial sampling errors which are one to two orders of magnitude less than the variability in the data as is immediately apparent from the points, the lines and the binomial errors in Figure 1. With such over-dispersed data one must allow for the total variability and use the residual sum-of-squares, from the best fit model, to estimate the changes in the AIC which gives the values indicated by $\Delta A_2$ in Table 1. The correct conclusion is therefore that comparing the ∫SI°S model to the SI°S model provides evidence in support of the role of heterogeneity but this is contradicted by comparing the ∫SI°S model to the SIS model which contains neither heterogeneity nor super-infection. Since the SIS model is simpler and has fewer parameters than the ∫SI°S model and they do not differ significantly one should probably favour the SIS model.

In a subsequent paper, Filion *et al.*[3] criticised the biological assumptions in the paper by Smith *et al.*[1] They argued that, while the data can be explained on the assumption that there is heterogeneity in the susceptibility of children to infection, the data can equally well be explained on the assumption that susceptibility to infection decreases as a power function of the EIR. In this they are correct since the data do not have the power to distinguish between their model[3] and the previous model.[1]

## Conclusion

Because only 6% of the variability in the data is accounted for by the binomial sampling errors the remaining 94%, which is due to differences among the surveys including geographical, climatic, entomological, sampling and methodological differences in the various surveys, is not accounted for in the analysis. A better analysis of the data would have made it clear that these data provide only weak support for the inclusion of heterogeneity but one cannot reject the SIS model, which does not include heterogeneity or super-infection, in favour of the ∫SI°S model.

While the authors' conclusions concerning heterogeneity, parasite clearance rates and immunity may be correct they are not supported by the evidence presented in their paper. Other ways need to be found to establish these key biological parameters in malaria transmission.

## Acknowledgements

We thank Prof. David Smith for giving us access to the original data.